\documentclass[a4paper,10pt,prl,twocolumn]{revtex4}
\usepackage[dvips]{graphics,color}
\usepackage{amsmath}
\usepackage{amsfonts}

\usepackage{amssymb}
\usepackage{amstext}
\usepackage[sort&compress]{natbib}
\usepackage{graphics,graphicx}

\newcommand{\be}{\begin{equation}}
\newcommand{\ee}{\end{equation}}
\newcommand{\bea}{\begin{eqnarray}}
\newcommand{\eea}{\end{eqnarray}}
\newcommand{\ket}[1]{\left\vert #1    \right\rangle }
\newcommand{\bra}[1]{\left\langle   #1  \right\vert}
\newcommand{\ave}[1]{\left\langle #1   \right\rangle }

\begin{document}

\title{Robust and optimal laser cooling of trapped ions}

\author{J. Cerrillo}
\email{j.cerrillo@imperial.ac.uk}
\author{A. Retzker}%
\author{M. B. Plenio}
\affiliation{Institute for Mathematical
Sciences, Imperial College London, SW7 2PG, UK}
\affiliation{QOLS,
The Blackett Laboratory, Imperial College London, Prince Consort
Rd., SW7 2BW, UK}

\begin{abstract}
We present a robust and fast laser cooling scheme suitable for trapped
atoms and ions. Based on quantum interference, generated by a special
laser configuration, it is able to rapidly cool the system such that
the final phonon occupation vanishes to zeroth order in the Lamb-Dicke
parameter in contrast to existing cooling schemes. Furthermore, it is
robust under conditions of fluctuating laser intensity and
frequency, thus making it a viable candidate for experimental
applications.
\end{abstract}

\maketitle

{\em Introduction ---} Laser cooling is a crucial ingredient for
probing quantum properties of matter \cite{Chu1998,Cohen1998,Phillips1998}.
It is a key factor in a wide variety of experiments ranging from
Bose-Einstein condensates and quantum computing to quantum simulation
with atoms and ions. Variants of cooling schemes range from Doppler
cooling for free particles \cite{Hansch1975} and its partner, side
band cooling for bound particles \cite{Wineland1975,Wineland1978},
to dark state cooling schemes for free \cite{Aspect1988} and bound
particles \cite{ZollerDark} relying on quantum interference which
arises thanks to their non-trivial internal electronic structures.

At present, sideband cooling is the method of choice for trapped
ions. It is a necessary requirement for efficient cooling that
motional sidebands with frequency $\nu$ can be resolved, i.e.,
the linewidth of the optical transition $\gamma\ll\nu$. Cooling is
then achieved by the red sideband transition which excites the atom
while at the same time annihilating a phonon to ensure energy
conservation. This transition rate must be higher than that on the
carrier and blue sideband transitions that heat the system either
through recoil after spontaneous decay (carrier) or coherent
generation of a phonon (blue sideband). The selection of the red
transition is ensured by the rotating wave approximation, i.e. energy
conservation, as long as the Rabi frequency $\Omega$ of the laser
satisfies $\Omega\ll\nu$. One method to suppress the carrier and
blue sideband transitions employs destructive interference exhibited
for example in dark states \cite{ZollerDark}. For this reason,
Electromagnetically Induced Transparency (EIT) \cite{Imamoglu2005}
has become an inspiration
for a variety of proposed laser cooling schemes for trapped ions
such as EIT cooling \cite{Morigi2000} and Stark-shift cooling
\cite{Retzker2007}. In EIT cooling, interference eliminates the
carrier transition to improve cooling performance while in the
Stark-shift scheme this is achieved in a rotated basis in a suitable
interaction picture \cite{Jonathan2000}.

In EIT cooling \cite{Morigi2000}
the existence of a dark state of a three level system allows
final temperatures below $\hbar\gamma/k_B$. This is achieved
in a three level scheme subject to Raman lasers with
strong blue single-photon detuning that couple both the
ground state and a meta-stable state to an excited dissipative
state. Among the dressed states of the system there is one
dark state that cancels the carrier transition. Well chosen
parameters, can center the red sideband transition under a
peak of the Fano-like absorption spectrum, while constraining
the blue sideband to a region with negligible absorption, thus
achieving low final temperatures. The final state of the system
is then
\begin{equation}
        \rho^{(EIT)}=|dark\rangle\langle dark|\otimes\sum_na_n|n\rangle\langle n|+o(\eta^2),
\label{nonpure}\end{equation}
where $\ket {dark}$ is the internal level steady state and $\ket n$ are the number states of the external degrees of freedom (dof), with a final mean phonon number of order $\left(\gamma/4\vert \Delta \vert\right)^2$
\cite{Morigi2000}.


In Stark-shift cooling \cite{Retzker2007} on the other hand one
laser drives transitions between the ground and a meta-stable state
and another two resonant
Raman lasers couple a superposition of both to the excited state.
This first laser generates Rabi oscillations between the dark state
and the orthogonal bright state. If the coupling is correctly tuned,
the oscillations will also involve neighboring mechanical levels so
that states $\ket {dark} \ket n$ and $\ket {bright} \ket {n-1}$ are
coupled while carrier transitions are
eliminated. As the EIT scheme, Stark-shift cooling achieves a final
temperature that is in leading order independent of the Lamb-Dicke
parameter $\eta$.

It is now natural to ask whether these schemes can be combined to
reach vanishing temperature in leading order in the Lamb-Dicke
parameter.
As this work demonstrates, this is indeed possible. We present
here a novel cooling scheme, a judicious combination of
EIT and Stark-shift cooling, which share the same dark state. This
in turn allows for both the carrier and the blue sideband transitions,
and hence heating, to be suppressed by interference. As a result
the final temperature, to zeroth order in the Lamb-Dicke parameter,
vanishes. To the best of our knowledge this is the first scheme
that is able to achieve this for trapped particles.


That the combination of EIT and Stark-shift cooling is successful,
may seem surprising. A classical analysis would lead one to expect
that the efficiency of the combined scheme to be the arithmetical
average of the individual ones. It is the quantum nature of the
interaction however that makes them interfere constructively in such
a way as to outperform its constituent schemes.
The quantum interference can be designed to completely cancel
the two leading heating processes in the system resulting in the state
\begin{equation}
\rho^{(robust)}=|dark\rangle\langle dark|\otimes|0\rangle\langle 0|+o(\eta^2)
\label{steadyrho}
\end{equation}
with a final temperature that vanishes in leading order in the
Lamb-Dicke parameter.
Remarkably we will also demonstrate that the scheme also exhibits
a significant improvement with regards to the stability under
fluctuating laser intensities. This implies that it remains robust
under general experimental conditions, thus assuring its feasibility.

\begin{figure}
\begin{center}
\includegraphics[width=0.3\textwidth]{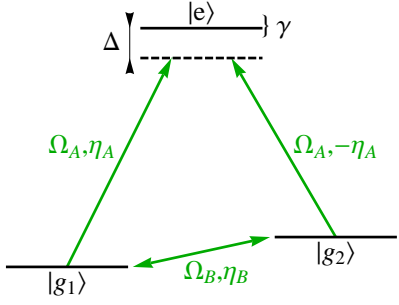}
\end{center}
\caption{A three level structure made up of $|g_1\rangle$ and
$|g_2\rangle$ and an excited dissipative state $|e\rangle$ which
decays at rate $\gamma$. The ground levels are coupled to the
excited state by a pair of Raman beams under a detuning $\Delta$
whose Rabi frequencies are $\Omega_A$ and mechanically interact
with the ion with opposite Lamb-Dicke parameters $\eta_A$. The
ground levels are directly coupled by an effective Rabi
frequency $\Omega_B$ and a Lamb-Dicke parameter $\eta_B$.}
\label{levels}
\end{figure}

{\em Description of robust cooling --}
The combined scheme, as presented in fig. \ref{levels}, involves a
trapped ion with an internal electronic structure made up of a ground
state and a metastable state $|g_1\rangle$ and $|g_2\rangle$ and an
excited dissipative state $|e\rangle$. For simplicity, we will refer
to the coupling between the lower states $|g_1\rangle$ and $|g_2\rangle$
as ``SSh coupling'' and between the g-manifold and the excited state
$|e\rangle$ as ``EIT couplings''. A harmonic well models the trap
potential. It is characterized by its equally spaced levels $\ket n$
representing the Fock state of n phonons and the creation and
annihilation operators $b$ and $b^\dagger$. Stark-shift and EIT
coolings can be regarded as particular instances of this scheme
when $\eta_{A}\rightarrow 0$ and $\Omega_{B}\rightarrow 0$
respectively.

The quantum theory of laser cooling of trapped particles was
developed in \cite{lindberg1986,Javanainen1981c,cirac1992} and we
will follow the notation in \cite{cirac1992}. Expanding the
Hamiltonian up to first order in the Lamb-Dicke parameters, it
can be split into three contributions, the trap potential and
the internal dof:
\be
        H_0=\nu b^\dagger b - \sum_{i} \delta_i \ket{g_i} \bra{g_i},
\label{free}
\ee
where are $\delta_i$'s are the detunings with respect to the lasers.
The interaction of the laser beams with the ion due to the EIT part:
\be
H_{EIT}= \Omega_A\left( \sigma_x^{g_1,e}+\sigma_x^{g_1,e}\right)+ \left(\sigma_y^{g_1,e}-\sigma_y^{g_2,e}\right)(b+b^\dagger)
\label{eit}
\ee
and the interaction of the laser beams with the ion due to the Stark-shift part:
\be
H_{SSh}= \Omega_B \sigma_x^{g_1,g_2} \eta_A \Omega_A + \eta_B \Omega_B \sigma_y^{g_1,g_2}(b+b^\dagger)
\label{ssh}
\ee
where
$\sigma_x^{(m,n)}=\left(|m\rangle\langle n|+h.c.\right)$ and
$\sigma_y^{(m,n)}=\left(i|m\rangle\langle n|+h.c.\right)$.

The physics of the scheme can best be understood by extending
the analysis of the steady state Eq.(\ref{steadyrho}) to the next
order in the Lamb-Dicke parameter. By inspection of the Hamiltonian
one can consider the following non-normalized pure state:
\begin{equation}
        |\Psi\rangle=|g_1-g_2\rangle|0\rangle
        -i\eta_B|g_1+g_2\rangle|1\rangle+o(\eta^2)
\label{steadystate}
\end{equation}
Under the effect of the EIT couplings (Eq. \ref{eit}) both terms
of the superposition interfere destructively and as a result
$|\Psi\rangle$ remains invariant. However $|\Psi\rangle$ is not
invariant under the free Hamiltonian (Eq. (\ref{free})) which introduces 
a phase shift between the two components.
For a suitable tuned parameters, the  Stark shift coupling part
(Eq. \ref{ssh}) can cancel the effect of the free Hamiltonian 
and as a consequence $|\Psi\rangle$ will be a dark state of
the Liouvillian as it does not suffer any spontaneous emission
losses either. This is achieved when
\begin{equation}
\frac{\eta_B}{\eta_A}=\left(\frac{\nu}{\Omega_B}+2\right).
\label{resonance}
\end{equation}
From an experimental point of view, it is worth noting that this
resonance condition is characterized by the quotient of the
Lamb-Dicke parameters. These can be set up at the beginning
of the experiment to a high precision.


Losses from the excited level are incorporated in the master
equation:
\begin{equation}
\frac{d\rho}{dt}=-i\hbar\left[H_{0}+H_{EIT}+H_{SSh},\rho\right]+\mathcal{L}^d\rho
\end{equation}
where $\mathcal{L}^d$ contains the spontaneous emission of the excited level:
\be
\mathcal{L}^d = \sum_{i=g1,g2} \gamma_{e,i} 2 \sigma_{i,e}\overline{\rho_{e,i}}\sigma_{e,i} -
\rho\sigma_{e,e}  - \sigma_{e,e} \rho
\ee
where $\sigma_{j,k}= \ket{j}\bra{k}$ and $\overline{\rho} = \frac{1}{2}\int_{-1}^1 dx e^{ikx} \rho e^{-ikx}.$

\begin{figure}
\includegraphics[width=0.5\textwidth]{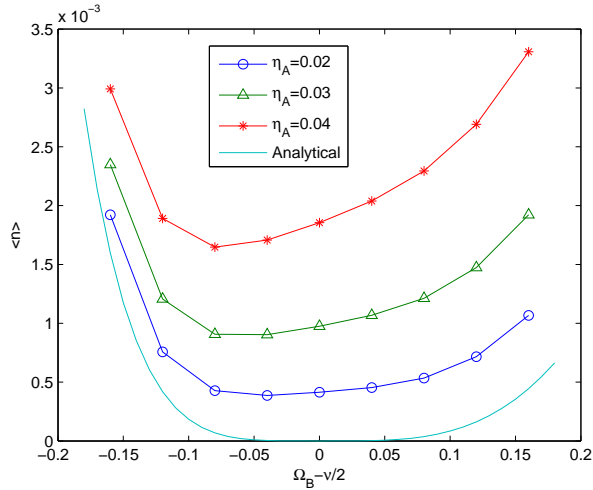}
\caption{The deviation of the analytical from the exact numerical
results in the final population is plotted versus the deviation
in $\Omega_B$ (in units of trap frequency) from the optimal operating
point for different Lamb-Dicke parameters. }
\label{numer}
\end{figure}

After expansion of the rest of the terms up to second order in
the Lamb-Dicke parameter and adiabatic elimination of the internal
dof we find the rate equation:
\begin{equation}
\begin{split}
\frac{d \rho^{n,n}_{ext}}{dt}=[(n+1)(A_-&\rho^{n+1,n+1}_{ext}-A_+\rho^{n,n}_{ext})\\&+n(A_+\rho^{n-1,n-1}_{ext}-A_-\rho^{n,n}_{ext})]
\end{split}
\end{equation}
In the spirit of \cite{cirac1992} the rates $A_\pm$ can be expressed
\begin{equation}
A_\pm=2 Re[D+S(\mp\nu)]
\end{equation}
where $D$ is the diffusion coefficient due to spontaneous emission
from the excited atomic states. Here $D=0$ as the population
of the excited states vanishes due to the dark state nature of
the final state. $S(\nu)$ is the fluctuation spectrum of Heisenberg
operator $F(t)$:
\begin{equation}
        S(\nu)=\frac{1}{2M\nu}\int^\infty_0dt e^{i\nu t}
        \langle F(t) F(0)\rangle
\end{equation}
where $F=F_{EIT}+F_{SSh}$ and $F_{EIT}=\sigma_y^{g_1 e}-\sigma_y^{g_2 e}$
and $F_{SSh} = \eta_B\Omega_B\sigma_y^{g_1g_2}$ are the part in the
interaction Hamiltonian Eq. \ref{eit}
and Eq. \ref{ssh} that multiply $b+b^\dagger$. The average can
be calculated using the quantum regression
theorem. We may decompose the operator $F$ into components of EIT and Stark-shift
so that the overall heating rate can then be split into three parts: the EIT part
$A_+^{EIT}$, the Stark-shift part $A_+^{SSh}$ and interaction
between EIT and the Stark-shift part
$A_+^{int}$ for the remaining cases. Their results are:
$A_+^{EIT}=(\eta_A (\nu + 2\Omega_B))^2/\mathcal{D}$,
$A_+^{SSh}=(\eta_B \Omega_B)^2/\mathcal{D}$ and
$A_+^{int}=-2 \eta_A\eta_B(\nu+2\Omega_B)\Omega_B/\mathcal{D}$
where
$\frac{1}{\mathcal{D}}=\frac{ 2\Omega_A^2\Gamma}{\Gamma^2(\nu +2 \Omega_B)^2 + ( -2\Omega_A^2 + (\nu + 2\Omega_B) (\Delta + \nu + \Omega_B))^2}$. Then
\begin{equation}
        A_+=[\eta_A (\nu + 2\Omega_B)-\eta_B \Omega_B]^2/\mathcal{D}
\end{equation}
which will be identically zero for the condition in
Eq.(\ref{resonance}). This emphasizes the role of the interaction
between the EIT and Stark-shift cooling in order to assure the
ground state is populated with probability $1$. The mean occupation
number $n$ is given by $\frac{A_+}{A_--A_+}$, thus yielding
$\langle n\rangle = 0$.
Fig. \ref{numer} shows the numerical results of the approach to the $\ave n =0$ point as the Lamb Dicke parameter goes to zero.

\begin{figure}
    \includegraphics[width=0.45\textwidth,height=0.17\textheight]{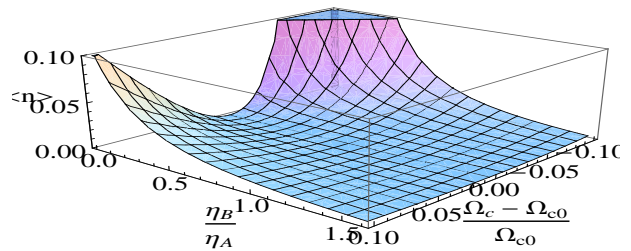}
   \caption{
Expected value of occupation $\langle n\rangle$ as a function of
the Lamb-Dicke parameter quotient and of variations around the
optimal Rabi frequency $\Omega_B$, where it is shown how variations
have a worse effect as $\eta_{A}$ vanishes, that is, as the
Stark-shift cooling regime is approached. }
\label{fig:robust}
\end{figure}

{\em Robustness --- }
The constructive interference between EIT and Stark-shift contribution
is also crucial for understanding the robustness of the scheme
under fluctuating parameters. If the Rabi-frequencies deviate
from Eq.(\ref{resonance}) by, $\Delta \Omega_{A/B}$, the final
populations as
\begin{equation}
        \langle n\rangle\propto (\Delta\Omega_A)^4(\Delta\Omega_B)^2
\end{equation}
instead of second order as is usually the case.
As is exemplified in fig. \ref{fig:robust}, under a given value of the
fluctuations of the laser intensities, the final mean occupation
decreases abruptly as one moves away of the Stark-shift only or
EIT only regimes. This guarantees promising performance under
real experimental conditions, overcoming the main drawback of
the previous dark-state cooling schemes.

{\em Rate ---} The interference structure of this scheme is also
essential for the high cooling rate $W = A_{-} - A_{+}$. For the
resonance condition eq. (\ref{resonance}) we find
\begin{equation}
        W=\frac{8 \Gamma \eta_A^2 \nu^2 \Omega_A^2}{\left(2\Omega_A^2
        + (\nu -2 \Omega_B) \left(\Delta - \nu + \Omega_B\right)\right)^2
        +\Gamma^2 (\nu - 2\Omega_B)^2}.
\label{rate}
\end{equation}
$\Omega_B$ is the only Rabi frequency involved in the condition
Eq.(\ref{resonance}). As a function of $\Omega_B$, Eq.(\ref{rate})
takes the approximate shape of a squared Lorentzian, with a peak
close to $\Omega_B=\nu/2$ at which point the cooling rate expression
reduces to
\begin{equation}
        W = \frac{\Gamma \eta_B^2 \nu^2}{ 8\Omega_A^2}.
\label{optimalrate}
\end{equation}
This is also an optimal point for Stark-shift cooling
\cite{Retzker2007} but it should be noted that the cooling rate of
Robust cooling is slightly higher than that of Stark-shift cooling.
Since the internal dynamics should be much faster than the external one for the perturbation approach to work the analytic result is not valid  for  $\Omega_A < \eta_A \nu$, which means that the cooling can be as fast as 1 order of magnitude less than the trap frequency.
Since the final state (eq. \ref{steadystate}) is a pure state up to second order in the Lamb-Dicke parameter there is a simple unitary rotation such that the final state has vanishing number of phonons to fourth order in the Lamb Dicke parameter.

{\em Implementation ---}
The way the effective $\Omega_B$ and $\eta_B$ couplings can be
physically implemented is not unique. One way is to use two lasers
to create the EIT cooling part and to use magnetic gradients
\cite{Mintert2001} for the Stark-shift part. In this system the magnetic gradients create a coupling of the following type: $\lambda \sigma_z (b+b^\dagger)$, where $\lambda$ is proportional to the magnetic gradients and the TLS is driven using a microwave: $\Omega_d \sigma_x \cos {\omega_d t}$, where $\Omega_d$ corresponds to the Rabi frequency of the driving and $\omega_d$ to the driving frequency. After a polaron transformation the resulting Hamiltonian is exactly as in Eq. \ref{ssh} when the Rabi frequency is replaced by $\Omega_d$ and the Lamb Dicke parameter is replaced by $\frac{\lambda}{\nu}$.

This scheme can be especially useful to cool nano scale
resonators, by using the setup described in \cite{Rabl2009}.
In this setup an NV center is coupled to a diamond cantilever,
the coupling is achieved by magnetic gradients resulting in
the same Hamiltonian as described above. In cantilevers the
speed of cooling is very important due to the finite $Q$ value,
which is an important factor limiting the achievable final
temperatures at present. The high cooling rate achieved
by the described scheme will result in lower final temperatures
bringing us closer to the goal of achieving the quantum regime
in cantilever systems.

Alternatively, one can also use Raman beams with large single-photon
detuning to couple levels $|g_1\rangle$ and $|g_2\rangle$. By
adiabatically eliminating the upper level the relationships
between our effective parameters $\Omega_B$ and $\eta_B$ and
the physical values $\Omega_{p}$ and $\eta_{p}$ are found to be
$\Omega_B=\Omega_{p}^2\left[\frac{1}{\Delta}
-\frac{\Delta}{\Delta^2-\nu^2}\eta_{p}^2(2n+1)\right]
$ and
$\Omega_B\eta_B=\Omega_{p}^2\eta_{p}
\frac{2\Delta^2-\nu^2}{\Delta(\Delta^2-\nu^2)}.$
Neglecting the $\eta^2$ correction in the first expression we
find
$\eta_B=\eta_{phys}\frac{2\Delta^2-\nu^2}{\Delta^2-\nu^2}$
which yields $\eta_B=2\eta_{p}$ for sufficiently large detunings.

If we chose the optimal point for the cooling rate and
fluctuations $\Omega_B=\frac{\nu}{2}$, the condition becomes
$\eta_{p}/\eta_A=4$. This can be achieved for a layout where
beam B is colinear to the trap axis and the beam A is $60^o$
away from the axis.
Even if the optimal situation is for an angle separation of $60^o$, the robustness of the scheme ensures excellent performance for any geometrical configuration. Proposed experimental layouts   \footnote{personal communication with K. Singer and T. Sch\"atz.} use vacuum chambers with windows at $22.5^o$ and/or  $45^o$ from the trap axis, which generally allow for an angle range of about $\pm10^o$. Taking $45^o$ as an operating value, the Lamb-Dicke parameter quotient becomes $\eta_B/\eta_A=2\sqrt{2}$. Different optimal set of parameters can be obtained for this situation depending on whether the cooling rate or the final temperature want to be optimized. For the latter, condition Eq.(\ref{resonance}) has to be observed, $\Omega_A=0.6\nu$ and $\Delta\simeq 0$. This will assure an extremely stable cooling rate which is generally 2 orders of magnitude smaller than EIT cooling but still better than sideband cooling. This final result can be improved depending on the particular values of the transition linewidth $\Gamma$. On the contrary, if the cooling rate is to be enhanced, condition Eq.(\ref{resonance}) won't be satisfied. In particular, for $\Omega_A\simeq 0.4\nu$, $\Omega_B\simeq 0.45\nu$ with $5\%$ fluctuation and $\Delta\simeq -2\nu$ the population can still be as low as $10^{-3}$ while having a cooling rate several orders of magnitude over that of EIT cooling. Taking into account the fact that angles up to $55^o$ are accessible the cooling rate can still be improved by up to two orders of magnitude.


{\em Multi mode cooling --- } Finally, the cooling scheme has also
been tested for an ion chain using Monte Carlo simulation
\cite{MolmerJumps,PlenioJumps}. The robustness of the scheme implies
a wide range of operational Rabi frequencies or, in a different
perspective, a wide range of trap frequencies for a given Rabi
frequency.  Thus, a particular central mode frequency can be
addressed so that also the neighboring modes benefit from the
cooling. In a multi-mode environment with up to 3 ions promising
results have been obtained and a detailed study will be presented
elsewhere.

{\em Conclusion ---}  To conclude, we have introduced a cooling scheme that cools to zero temperature without any corrections in zeroth order in the Lamb Dicke parameter.
Beyond the academic interest of proving the existence of such a scheme, its robustness makes it extremely attractive for experimental realizations.

{\em Acknowledgment --- } J. C. acknowledges support from the
AXA Research Fund. A.R. acknowledges the support of EPSRC
project number EP/E045049/1 and MBP acknowledges support
from the Royal Society and the EU STREP project HIP.




\begin{thebibliography}{19}
\expandafter\ifx\csname natexlab\endcsname\relax\def\natexlab#1{#1}\fi
\expandafter\ifx\csname bibnamefont\endcsname\relax
  \def\bibnamefont#1{#1}\fi
\expandafter\ifx\csname bibfnamefont\endcsname\relax
  \def\bibfnamefont#1{#1}\fi
\expandafter\ifx\csname citenamefont\endcsname\relax
  \def\citenamefont#1{#1}\fi
\expandafter\ifx\csname url\endcsname\relax
  \def\url#1{\texttt{#1}}\fi
\expandafter\ifx\csname urlprefix\endcsname\relax\def\urlprefix{URL }\fi
\providecommand{\bibinfo}[2]{#2}
\providecommand{\eprint}[2][]{\url{#2}}

\bibitem[{\citenamefont{Chu}(1998)}]{Chu1998}
\bibinfo{author}{\bibfnamefont{S.}~\bibnamefont{Chu}}, \bibinfo{journal}{Rev.
  Mod. Phys.} \textbf{\bibinfo{volume}{70}}, \bibinfo{pages}{685}
  (\bibinfo{year}{1998}).

\bibitem[{\citenamefont{Cohen-Tannoudji}(1998)}]{Cohen1998}
\bibinfo{author}{\bibfnamefont{C.~N.} \bibnamefont{Cohen-Tannoudji}},
  \bibinfo{journal}{Rev. Mod. Phys.} \textbf{\bibinfo{volume}{70}},
  \bibinfo{pages}{707} (\bibinfo{year}{1998}).

\bibitem[{\citenamefont{Phillips}(1998)}]{Phillips1998}
\bibinfo{author}{\bibfnamefont{W.~D.} \bibnamefont{Phillips}},
  \bibinfo{journal}{Rev. Mod. Phys.} \textbf{\bibinfo{volume}{70}},
  \bibinfo{pages}{721} (\bibinfo{year}{1998}).

\bibitem[{\citenamefont{Hansch and Schawlow}(1975)}]{Hansch1975}
\bibinfo{author}{\bibfnamefont{T.}~\bibnamefont{Hansch}} \bibnamefont{and}
  \bibinfo{author}{\bibfnamefont{A.}~\bibnamefont{Schawlow}},
  \bibinfo{journal}{Opt. Commun} \textbf{\bibinfo{volume}{13}},
  \bibinfo{pages}{68} (\bibinfo{year}{1975}).

\bibitem[{\citenamefont{Wineland and Dehmelt}(1975)}]{Wineland1975}
\bibinfo{author}{\bibfnamefont{D.}~\bibnamefont{Wineland}} \bibnamefont{and}
  \bibinfo{author}{\bibfnamefont{H.}~\bibnamefont{Dehmelt}},
  \bibinfo{journal}{Bull. Am. Phys. Soc} \textbf{\bibinfo{volume}{20}},
  \bibinfo{pages}{637} (\bibinfo{year}{1975}).

\bibitem[{\citenamefont{Wineland et~al.}(1978)\citenamefont{Wineland,
  Drullinger, and Walls}}]{Wineland1978}
\bibinfo{author}{\bibfnamefont{D.~J.} \bibnamefont{Wineland}},
  \bibinfo{author}{\bibfnamefont{R.~E.} \bibnamefont{Drullinger}},
  \bibnamefont{and} \bibinfo{author}{\bibfnamefont{F.~L.} \bibnamefont{Walls}},
  \bibinfo{journal}{Phys. Rev. Lett.} \textbf{\bibinfo{volume}{40}},
  \bibinfo{pages}{1639} (\bibinfo{year}{1978}).

\bibitem[{\citenamefont{Aspect et~al.}(1988)\citenamefont{Aspect, Arimondo,
  Kaiser, Vansteenkiste, and Cohen-Tannoudji}}]{Aspect1988}
\bibinfo{author}{\bibfnamefont{A.}~\bibnamefont{Aspect}},
  \bibinfo{author}{\bibfnamefont{E.}~\bibnamefont{Arimondo}},
  \bibinfo{author}{\bibfnamefont{R.}~\bibnamefont{Kaiser}},
  \bibinfo{author}{\bibfnamefont{N.}~\bibnamefont{Vansteenkiste}},
  \bibnamefont{and}
  \bibinfo{author}{\bibfnamefont{C.}~\bibnamefont{Cohen-Tannoudji}},
  \bibinfo{journal}{Phys. Rev. Lett.} \textbf{\bibinfo{volume}{61}},
  \bibinfo{pages}{826} (\bibinfo{year}{1988}).

\bibitem[{\citenamefont{Dum et~al.}(1994)\citenamefont{Dum, Marte, Pellizzari,
  and Zoller}}]{ZollerDark}
\bibinfo{author}{\bibfnamefont{R.}~\bibnamefont{Dum}},
  \bibinfo{author}{\bibfnamefont{P.}~\bibnamefont{Marte}},
  \bibinfo{author}{\bibfnamefont{T.}~\bibnamefont{Pellizzari}},
  \bibnamefont{and} \bibinfo{author}{\bibfnamefont{P.}~\bibnamefont{Zoller}},
  \bibinfo{journal}{Phys. Rev. Lett.} \textbf{\bibinfo{volume}{73}},
  \bibinfo{pages}{2829} (\bibinfo{year}{1994}).

\bibitem[{\citenamefont{Fleischhauer et~al.}(2005)\citenamefont{Fleischhauer,
  Imamoglu, and Marangos}}]{Imamoglu2005}
\bibinfo{author}{\bibfnamefont{M.}~\bibnamefont{Fleischhauer}},
  \bibinfo{author}{\bibfnamefont{A.}~\bibnamefont{Imamoglu}}, \bibnamefont{and}
  \bibinfo{author}{\bibfnamefont{J.~P.} \bibnamefont{Marangos}},
  \bibinfo{journal}{Rev. Mod. Phys.} \textbf{\bibinfo{volume}{77}},
  \bibinfo{pages}{633} (\bibinfo{year}{2005}).

\bibitem[{\citenamefont{Morigi et~al.}(2000)\citenamefont{Morigi, Eschner, and
  Keitel}}]{Morigi2000}
\bibinfo{author}{\bibfnamefont{G.}~\bibnamefont{Morigi}},
  \bibinfo{author}{\bibfnamefont{J.}~\bibnamefont{Eschner}}, \bibnamefont{and}
  \bibinfo{author}{\bibfnamefont{C.~H.} \bibnamefont{Keitel}},
  \bibinfo{journal}{Phys. Rev. Lett.} \textbf{\bibinfo{volume}{85}},
  \bibinfo{pages}{4458} (\bibinfo{year}{2000}).

\bibitem[{\citenamefont{Retzker and Plenio}(2007)}]{Retzker2007}
\bibinfo{author}{\bibfnamefont{A.}~\bibnamefont{Retzker}} \bibnamefont{and}
  \bibinfo{author}{\bibfnamefont{M.~B.} \bibnamefont{Plenio}},
  \bibinfo{journal}{New J. of Phys.} \textbf{\bibinfo{volume}{9}},
  \bibinfo{pages}{279} (\bibinfo{year}{2007}).

\bibitem[{\citenamefont{Jonathan et~al.}(2000)\citenamefont{Jonathan, Plenio,
  and Knight}}]{Jonathan2000}
\bibinfo{author}{\bibfnamefont{D.}~\bibnamefont{Jonathan}},
  \bibinfo{author}{\bibfnamefont{M.~B.} \bibnamefont{Plenio}},
  \bibnamefont{and} \bibinfo{author}{\bibfnamefont{P.~L.}
  \bibnamefont{Knight}}, \bibinfo{journal}{Phys. Rev. A}
  \textbf{\bibinfo{volume}{62}}, \bibinfo{pages}{042307}
  (\bibinfo{year}{2000}).

\bibitem[{\citenamefont{Lindberg and Javanainen}(1986)}]{lindberg1986}
\bibinfo{author}{\bibfnamefont{M.}~\bibnamefont{Lindberg}} \bibnamefont{and}
  \bibinfo{author}{\bibfnamefont{J.}~\bibnamefont{Javanainen}},
  \bibinfo{journal}{J. Opt. Soc. B} \textbf{\bibinfo{volume}{3}},
  \bibinfo{pages}{1008} (\bibinfo{year}{1986}).

\bibitem[{\citenamefont{Javanainen and Stenholm}(1981)}]{Javanainen1981c}
\bibinfo{author}{\bibfnamefont{J.}~\bibnamefont{Javanainen}} \bibnamefont{and}
  \bibinfo{author}{\bibfnamefont{S.}~\bibnamefont{Stenholm}},
  \bibinfo{journal}{Applied Physics A} \textbf{\bibinfo{volume}{24}},
  \bibinfo{pages}{151} (\bibinfo{year}{1981}).

\bibitem[{\citenamefont{Cirac et~al.}(1992)\citenamefont{Cirac, Blatt, and
  Zoller}}]{cirac1992}
\bibinfo{author}{\bibfnamefont{J.~I.} \bibnamefont{Cirac}},
  \bibinfo{author}{\bibfnamefont{R.}~\bibnamefont{Blatt}}, \bibnamefont{and}
  \bibinfo{author}{\bibfnamefont{P.}~\bibnamefont{Zoller}},
  \bibinfo{journal}{Phys. Rev. A} \textbf{\bibinfo{volume}{46}},
  \bibinfo{pages}{2668} (\bibinfo{year}{1992}).

\bibitem[{\citenamefont{Mintert and Wunderlich}(2001)}]{Mintert2001}
\bibinfo{author}{\bibfnamefont{F.}~\bibnamefont{Mintert}} \bibnamefont{and}
  \bibinfo{author}{\bibfnamefont{C.}~\bibnamefont{Wunderlich}},
  \bibinfo{journal}{Phys. Rev. Lett.} \textbf{\bibinfo{volume}{87}},
  \bibinfo{pages}{257904} (\bibinfo{year}{2001}).

\bibitem[{\citenamefont{Rabl et~al.}(2009)\citenamefont{Rabl, Cappellaro, Dutt,
  Jiang, Maze, and Lukin}}]{Rabl2009}
\bibinfo{author}{\bibfnamefont{P.}~\bibnamefont{Rabl}},
  \bibinfo{author}{\bibfnamefont{P.}~\bibnamefont{Cappellaro}},
  \bibinfo{author}{\bibfnamefont{M.~V.~G.} \bibnamefont{Dutt}},
  \bibinfo{author}{\bibfnamefont{L.}~\bibnamefont{Jiang}},
  \bibinfo{author}{\bibfnamefont{J.~R.} \bibnamefont{Maze}}, \bibnamefont{and}
  \bibinfo{author}{\bibfnamefont{M.~D.} \bibnamefont{Lukin}},
  \bibinfo{journal}{Phys. Rev. B} \textbf{\bibinfo{volume}{79}},
  \bibinfo{pages}{041302} (\bibinfo{year}{2009}).

\bibitem[{\citenamefont{Molmer et~al.}(1993)\citenamefont{Molmer, Castin, and
  Dalibard}}]{MolmerJumps}
\bibinfo{author}{\bibfnamefont{K.}~\bibnamefont{Molmer}},
  \bibinfo{author}{\bibfnamefont{Y.}~\bibnamefont{Castin}}, \bibnamefont{and}
  \bibinfo{author}{\bibfnamefont{J.}~\bibnamefont{Dalibard}},
  \bibinfo{journal}{J. Opt. Soc. Am. B} \textbf{\bibinfo{volume}{10}},
  \bibinfo{pages}{524} (\bibinfo{year}{1993}).

\bibitem[{\citenamefont{Plenio and Knight}(1998)}]{PlenioJumps}
\bibinfo{author}{\bibfnamefont{M.~B.} \bibnamefont{Plenio}} \bibnamefont{and}
  \bibinfo{author}{\bibfnamefont{P.~L.} \bibnamefont{Knight}},
  \bibinfo{journal}{Rev. Mod. Phys.} \textbf{\bibinfo{volume}{70}},
  \bibinfo{pages}{101} (\bibinfo{year}{1998}).

\end{thebibliography}
\end{document}